\documentclass[12pt,preprint]{aastex}

\newcommand{\skipthis}[1]{}

\shorttitle{SMA Observations of L1551 IRS 5} 
\shortauthors{Takakuwa et al.}

\begin{document}


\title{Submillimeter Array Observations of L1551 IRS 5 in CS ($J$=7--6)}

\author{Shigehisa Takakuwa\altaffilmark{1,2}, Nagayoshi Ohashi\altaffilmark{3}, 
Paul T. P. Ho\altaffilmark{4}, Chunhua Qi\altaffilmark{4}, 
David Wilner\altaffilmark{4},\\Qizhou Zhang\altaffilmark{4}, 
Tyler L. Bourke\altaffilmark{1}, Naomi Hirano\altaffilmark{5}, Minho Choi\altaffilmark{5,6}, 
\& Ji Yang\altaffilmark{7}}

\altaffiltext{1}{Harvard-Smithsonian Center for Astrophysics, Submillimeter Array Project, 
645 North A'ohoku, Hilo, HI 96720, U.S.A.}
\altaffiltext{2}{e-mail: stakakuwa@sma.hawaii.edu}
\altaffiltext{3}{Academia Sinica Institute of Astronomy and Astrophysics,
645 North A'ohoku, Hilo, HI 96720, U.S.A.; nohashi@sma.hawaii.edu}
\altaffiltext{4}{Harvard-Smithsonian Center for Astrophysics, 60 Garden Street, 
Cambridge, MA 02138, U.S.A.}
\altaffiltext{5}{Academia Sinica Institute of Astronomy and Astrophysics,
P.O. Box 23-141, Taipei 106, Taiwan}
\altaffiltext{6}{Taeduk Radio Astronomy Observatory, 
Korea Astronomy Observatory, Hwaam 61-1, Yuseong, Daejeon, 305-348, Korea}
\altaffiltext{7}{Purple Mountain Observatory, 
Chinese Academy of Sciences, Nanjing 21008, China}

\begin{abstract}
We have imaged the circumstellar envelope around the 
binary protostar L1551 IRS 5 in CS ($J$=7--6) and 343 GHz continuum emission 
at $\sim$ 3$\arcsec$ resolution using the Submillimeter Array. 
The continuum emission shows an elongated structure 
($\sim220 \times 100$ AU) 
around the binary perpendicular to the axis of the associated radio jet. 
The CS emission extends over $\sim400$~AU, 
appears approximately circularly symmetric, and shows a velocity 
gradient from southeast (blueshifted) to northwest (redshifted). 
The direction of the velocity gradient is different from that observed 
in C$^{18}$O ($J$=1--0). This may be because rotation is more dominant in the CS
envelope than the C$^{18}$O envelope, in which both infall and rotation exist.
The CS emission may be divided into two velocity components: 
(1) a ``high" velocity disk-like structure surrounding the protostar, 
$\pm1.0-1.5$~km~s$^{-1}$ from the systemic velocity, and 
(2) a ``low" velocity structure, located southwest of the protostar,
$<1.0$~km~s$^{-1}$ from the systemic velocity.
The high-velocity component traces warm and dense gas with kinematics
consistent with rotation around the protostar.
The low-velocity component may arise from dense gas entrained in the outflow.
Alternatively, this component may trace 
infalling and rotating gas in an envelope with a vertical structure. 

\end{abstract}


\section{Introduction}


It is widely accepted that low-mass stars are formed through mass accretion
in protostellar envelopes \cite{and00,mye00}.  The dense and warm innermost
part of low-mass protostellar envelopes (radius $<$ 200 AU) are likely
sites of formation of protoplanetary disks \cite{spa95,bec96}.  However,
we have very little knowledge about this region because previous mm-wave
observations could not separate the warm and dense regions from the
overlying low-density ($\sim10^{4-5}$ cm$^{-3}$) and cold
($\sim$10 K) gas along the line of sight. On the other hand, submillimeter
molecular lines such as CS ($J$=7--6) trace higher densities ($>$
10$^{\rm 7}$ cm$^{\rm -3}$) and temperatures ($>$ 60 K)
\cite{gms95,spa95}, and submillimeter dust continuum emission is an
excellent tracer of disks around protostars \cite{oso03}. 

In this $Letter$, we describe results of CS ($J$=7--6) and 343 GHz
continuum observations of L1551 IRS 5 with the Submillimeter Array
(SMA)\footnote{The Submillimeter Array (SMA) is a joint project between the
Smithsonian Astrophysical Observatory and the Academia Sinica Institute of
Astronomy and Astrophysics, and is funded by the Smithsonian Institution
and the Academia Sinica.}.  L1551 IRS 5 is the brightest protostar
($L_{bol}$ $\sim30 L_{\odot}$; Keene \& Masson 1990) in Taurus ($D \sim140$
pc; Elias 1978). After the first discovery of the bipolar molecular outflow 
by Snell, Loren \& Plambeck (1980), Kaifu et~al.\ (1984) discovered a 0.1 pc scale rotating
gas structure around L1551 IRS 5 with the Nobeyama 45 m telescope.  The
elongated rotating protostellar envelope around L1551 IRS 5 was first
imaged by Sargent et~al.\ (1988) in C$^{18}$O ($J$=1--0) with the OVRO
mm-array.  Ohashi et~al.\ (1996), Saito et~al.\ (1996), and Momose et~al.\
(1998) have found evidence of infalling motion in the protostellar envelope
with the Nobeyama Millimeter Array (NMA). Millimeter interferometric
observations of dust continuum emission from L1551 IRS 5 revealed that
L1551 IRS 5 is a close ($\sim0\farcs3$) binary \cite{loo97,rod98}.  Lay
et~al.\ (1994) made the first submillimeter interferometric observations of
dust continuum emission from L1551 IRS 5 with the JCMT-CSO interferometer.
Single-dish observations of L1551 IRS 5 in submillimeter molecular lines
have revealed warm ($\geq$ 40 K) and dense ($\geq$ 10$^{6}$ cm$^{-3}$) gas
in the inner part of the envelope \cite{ful95,gms95,hog97,hog98}.  Our SMA
observations provide us with new information of the innermost part of the
protostellar envelope.

\section{Observations}

With the SMA, CS ($J$=7--6; 342.88295 GHz) and 343 GHz continuum
observations of L1551 IRS 5 were made on 2002 December 18 and 2003 March
13.  Details of the SMA are described by Ho, Moran, \& Lo (2004).  Table 1
summarizes the observational parameters.  We confirmed that the visibility
amplitudes of the continuum emission from L1551 IRS 5 at the two observing
periods were consistent within the noise level.  The overall flux
uncertainty was estimated to be about 20 $\%$.  Since the minimum baseline
length projected on the sky was 14.8 k$\lambda$, our observations were
insensitive to structures more extended than 11$\arcsec$ ($\sim$ 1500 AU)
at the 10\% level \cite{wil94}.

The raw visibility data were calibrated and flagged with MIR, and the
calibrated visibility data were Fourier-transformed and CLEANed with
MIRIAD.  We adopted robust = 0.5 weighting for the imaging, which seemed to
provide the best compromise between the sensitivity and the spatial
resolution.  The rms noise levels of the images were about twice of what is
expected, for reasons that are under investigation. 

\placetable{tlb-1}

\section{Results}

In Figure 1, we show the total integrated intensity map of the CS
($J$=7--6) emission along with the C$^{\rm 18}$O ($J$=1--0) map taken with
the NMA by Momose et~al.\ (1998).  While the C$^{\rm 18}$O map reveals a
structure of the protostellar envelope, elongated ($\sim$ 2400 AU $\times$
1000 AU in size) perpendicularly to the axis of the associated molecular
outflow \cite{uch87,gms88}, the CS emission shows a structure without clear
elongation.  From the elliptical Gaussian fitting, the size of this CS
condensation deconvolved by the synthesized beam was derived to be $\sim$
3\farcs4 $\times$ 2\farcs4 (P.A. 40$^{\circ}$), which corresponds
to the diameter of $\sim$ 400 AU at the distance of L1551 IRS 5. 
Comparison of the SMA flux to the total flux of the CSO CS ($J$=7--6)
spectrum toward L1551 IRS 5 suggests that $\sim11\%$ of the single-dish
flux is recovered by the SMA at $\sim3\arcsec$ resolution. 

In Figure 2, we show the intensity-weighted mean velocity map of L1551 IRS
5 in the CS emission. There is a clear velocity gradient from southeast to
northwest, which is perpendicular to the axis of the associated molecular
outflow (NE-SW; Uchida et~al.\ 1987, Moriarty-Schieven \& Snell 1988). This
direction of the velocity gradient is different from that observed in the
C$^{18}$O emission \cite{mom98}, e.g., the C$^{18}$O envelope shows a
velocity gradient from south to north near the central star. The reason of
this difference may be because the CS envelope is more dominated by
rotation as compared to the C$^{18}$O envelope, which is dominated by
infall as well as rotation.

\placefigure{f1}

In order to understand the velocity structure better, we made velocity
channel maps of the CS emission at different velocity ranges, that is, a
``high-velocity'' (5.31 - 5.85 km s$^{-1}$ and 7.82 - 8.35 km s$^{-1}$)
component and a ``low-velocity'' (6.39 - 6.74 km s$^{-1}$ and 6.92 - 7.28
km s$^{-1}$) component.  In Figure 3 we show the high-velocity and
low-velocity components separately, superposed on the 343 GHz dust
continuum emission and the VLA 3.5 cm radio jet image \cite{rod03}.  
The total 343 GHz continuum flux is 2.2 Jy with a deconvolved size of 
1\farcs6 $\times$ 0\farcs7 (220 $\times$ 100 AU) at P.A.\ $-$9.6\degr.
derived from the elliptical Gaussian fitting.
The flux and size are consistent with the
estimates from the JCMT-CSO interferometric study \cite{lay94}, and the
dust emission most probably represents the circumbinary disk and the
circumstellar disks associated with the protostars, although the
circumstellar disks are not spatially resolved.  The high-velocity
component in the CS emission spatially coincides with the dust continuum
distribution which is approximately perpendicular to the axis of the radio
jet \cite{rod03}.  On the other hand, there is a slight, but significant
spatial discrepancy between the low-velocity component and the dust
continuum emission, wherein the low-velocity component is located to the
southwest of the dust disk.  The peak of the redshifted low-velocity
component seems to lie along the axis of the radio jet.  We discuss these
different velocity components in more detail below. 

\placefigure{f2}

\placefigure{f3}

\section{Discussion}
\subsection{Physical Conditions of the CS Emitting Region}

Our CS ($J$=7--6) observations have revealed a
rather compact ($\sim400$ AU) structure in the inner part of the elongated
($\sim2000$ AU) protostellar envelope detected by mm-wave observations
\cite{oh96b,sai96,mom98} around L1551 IRS 5.  The critical gas density
traced by the CS line is expected to be $\geq$ 10$^{\rm 7}$ cm$^{\rm -3}$,
and the equivalent temperature of the $J$=7 energy level is 66 K.  This gas
density and temperature are much higher than those traced by the
millimeter-wave tracers, such as C$^{\rm 18}$O ($J$=1--0) ($\sim10^{4}$ 
cm$^{\rm -3}$; 5 K) \cite{sar88,mom98} or H$^{\rm 13}$CO$^{\rm +}$
($J$=1--0) ($\sim10^{5}$ cm$^{-3}$; 4 K) \cite{sai96}.  Indeed,
Moriarty-Schieven et~al.\ (1995) estimated gas density and temperature at
the center of the envelope in L1551 IRS 5 to be $\sim10^{7}$
cm$^{-3}$ and $\sim40$ K from their multi-transitional CS ($J$=3--2,
5--4, 7--6) observations with JCMT and CSO. Part of the difference in the
distribution between the CS ($J$=7--6) and C$^{\rm 18}$O ($J$=1--0)
emission shown in Figure 1 is likely to be due to the different physical
conditions traced by these two molecular lines.  The CS emission line
better traces the inner, denser and warmer part of the protostellar
envelope.

The compact shape of the CS ($J$=7--6) emission without elongation could be
due to the missing flux (only $\sim$ 11$\%$ recovered).  However, our
simple simulation indicated that, if the distribution of the CS ($J$=7--6)
emission were the same as that of the C$^{18}$O ($J$=1--0) emission, then
the present SMA observations would not detect any of the CS emission. This
suggests that there should be a compact structure of the CS emission as
well as an extended CS component with $\geq$ 11$\arcsec$ = 1500 AU in size.
We note that it is still a puzzle why the CS ($J$=7--6) emission which
traces denser and warmer gas can be so extended.  It seems to be difficult
to make gas temperature high enough only by the heating from the central
stars (e.g., Lay et~al.\ 1994).  There might be some external heating
sources such as shocks due to infalling and/or outflowing gas
\cite{nak00,ave96}.

\placefigure{f4}

\subsection{Origin of the High- and Low-Velocity Components}

As described in $\S$3, the CS ($J$=7--6) emission in L1551 IRS 5 consists
of a high-velocity disk-like component and a low-velocity component whose
peak locates at the southwest of the protostar.  In order to investigate
the origin of these velocity components in more detail, in Figure 4 we show
the position-velocity (P-V) diagram of the CS emission along the cut
perpendicular to the axis of the radio jet (see Figure 3).  The velocity
structure appears to be symmetric with respect to the velocity of 6.8 km
s$^{-1}$, and we adopt this velocity as a systemic velocity in this paper
(Note that this is significantly different from the systemic velocity of
C$^{\rm 18}$O ($J$=1--0) of 6.2 km s$^{-1}$; Momose et~al.\ 1998).
Solid color curves in Figure 4 indicate the Keplerian rotation velocity
($\propto$ $r^{-0.5}$) for a disk with an inclination angle of 65$^{\circ}$
\cite{mom98} and with a central stellar mass of 0.15 M$_{\odot}$ (red), 0.5
M$_{\odot}$ (green), and 1.2 M$_{\odot}$ (blue). The central stellar mass
of 0.15 M$_{\odot}$ was estimated by Momose et~al.\ (1998) from their
detailed kinematic analyses of the C$^{\rm 18}$O infalling envelope, and
that of 1.2 M$_{\odot}$ by Rodr\'{\i}guez et~al.\ (2003a) from the proper
motion analyses of the binary.  Dashed curves in Figure 4 show rotation
with angular momentum conserved, that is, $V_{rot} =
0.24(r/700{\rm AU})^{-1}~~{\rm km\ s^{-1}}$.
Momose et~al.\ (1998) argued that this rotation law can explain the observed
P-V diagram of the C$^{\rm 18}$O infalling envelope.  The high-velocity
component, with relative velocities of $\pm$ 1.0--1.5 km s$^{-1}$ in Figure
4, seems to be consistent with the Keplerian rotation with a central
stellar mass of 0.15 M$_{\odot}$ or the rotation with the constant angular
momentum, and is likely to be the rotating circumbinary disk at the
innermost of the envelope.  Unfortunately, it is difficult for us to
distinguish these two rotation curves to explain the kinematics of the
high-velocity component with the current sensitivity.  Keplerian rotation
curves with higher ($>$ 1.0 M$_{\odot}$) stellar masses show too high
velocities as compared to the CS P-V diagram.

In contrast, the low-velocity component ($\pm$0.5 km s$^{-1}$ from the
systemic velocity) seems not to follow these rotation curves.  The spatial
offset of the low-velocity component from the dust disk and the possible
coincidence with the radio jet (Figure 3) may suggest that the origin of the
low-velocity component is outflow-related.  Interestingly, at the
south-western tip of the radio jet where the peaks of the low-velocity
component locate, Bally, Feigelson, \& Reipurth (2003) have detected X-ray
emission with $Chandra$, which they attribute to shocks caused by the jet.
One interpretation therefore is that the low-velocity component is the
shocked warm and dense molecular material pushed by the radio jet.
However, one difficulty of this interpretation may be its ``low velocity''.
The projected separation between the peaks of the low-velocity component
and the continuum peak is $\sim$ 100 AU, and the line of sight velocities
of the low-velocity component is only $\sim$ a few $\times$ 0.1 km
s$^{-1}$. In order for the low-velocity component to escape from the
gravitational potential of the central star with a mass of 0.15 M$_{\odot}$
\cite{mom98}, the angle between the flow and the plane of sky must be less
than 10$^{\circ}$, while Uchida et~al.\ (1987) estimated the inclination
angle of 15$^{\circ}$ from their large-scale CO outflow map.

It is important to compare our P-V diagram with those calculated using a
model envelope with a vertical structure, because models of flared
envelopes having infall as well as rotation yield peaks at lower velocity
near the central star in their P-V diagrams cut along the major axis
\cite{mom98,hog01}.  The reason why these peaks at lower velocity appear
near the central star is because infalling motions in the near- and
far-side of the envelope are observed together with rotation in the same
line of sight. These peaks cannot be explained with rotation
curves calculated using spatially thin disk models, which is very similar
to our case shown in Figure 4. It is therefore possible that a model envelope
having vertical structures with infall and rotation motions may explain the
lower velocity component.

The effect of the missing flux prevents us from arriving at a clear
interpretation.  The lower-velocity component could be part of spatially
extended component, which can often be seen at lower velocities.  Further
observations with shorter spacings using the SMA should provide us with the
more obvious view and interpretation of these velocity structures of warm
and dense gas in the innermost part of the protostellar envelope. 

\acknowledgments 
We are grateful to M. Momose and M. Saito for fruitful 
discussions. We would like to thank all the SMA staff supporting this 
work. S.T. and T.B. are supported by a postdoctoral fellowship of the 
Smithsonian Astrophysical Observatory. 


\clearpage

\figcaption[f1.eps]{Total integrated intensity map of the CS ($J$=7--6)
emission of L1551 IRS 5 taken with the SMA (pseudo color and black contour)
with the C$^{18}$O ($J$=1--0) total intensity map taken with the NMA by
Momose et al. (1998) (light blue contour). Contour levels of the SMA
observations are from 3.0 Jy beam$^{-1}$ km s$^{-1}$ in steps of 2.0 Jy
beam$^{-1}$ km s$^{-1}$.  White crosses indicate the position of the binary
\cite{rod98}.  The SMA synthesized beam is shown at lower right, and the
NMA synthesized beam at lower left.  \label{f1}}

\figcaption[f2.eps]{CS ($J$=7--6) intensity-weighted mean velocity map in
L1551 IRS 5.  Contour levels are from 5.8 km s$^{-1}$ in steps of 0.2 km
s$^{-1}$.  Crosses indicate the position of the binary. There is clear
velocity gradient from southeast (blue) to northwest (red).  \label{f2}}

\figcaption[f3.eps]{CS ($J$=7--6) velocity channel maps of the
high-velocity (Left; Blue: 5.31 - 5.85 km s$^{-1}$, Red: 7.82 - 8.35 km
s$^{-1}$) and the low-velocity (Right; Blue: 6.39 - 6.74 km s$^{-1}$, Red:
6.92 - 7.28 km s$^{-1}$) components in L1551 IRS 5, superposed on the 343
GHz dust continuum map (gray scale) and the VLA 3.5 cm radio jet image
(white contour; Rodr\'{\i}guez et~al.\ 2003b).  Contour levels are from 1.8
Jy beam$^{-1}$ in steps of 0.9 Jy beam$^{-1}$.  Crosses indicate the
position of the binary.  Tilted black lines indicate the cut of the
position-velocity diagram shown in Figure 4.  \label{f3}}

\figcaption[f4.eps]{P-V diagram of the CS ($J$=7--6)
emission along the major axis (P.A. = 162$^{\circ}$; see Fig. 3) in L1551
IRS 5.  Contour levels are from 2.3 Jy beam$^{-1}$ in steps of 1.15 Jy
beam$^{-1}$.  Solid color curves indicate the Keplerian rotation velocity
($\propto$ $r^{-0.5}$) with a central stellar mass of 0.15 M$_{\odot}$
(red), 0.5 M$_{\odot}$ (green), and 1.2 M$_{\odot}$ (blue), and an
inclination angle of 65$^{\circ}$, while dashed curves indicate rotation
with angular momentum conservation as given by $V_{rot}$ in the text.  
\label{f4}}


\clearpage
\begin{deluxetable}{lcc}
\tablecaption{Parameters for the SMA Observations of L1551 IRS 5
  \label{tbl-1}} 
\tablewidth{0pt}
\tablehead{\colhead{Parameter} & \multicolumn{2}{c}{Value}\\
\cline{2-3}
\colhead{} & \colhead{2002 Dec 18} & \colhead{2003 Mar 13} }
\startdata
Number of Antennas& 3&  5 \\
\ \ \ Right Ascension (J2000) 
   & \multicolumn{2}{c}{04$^{\rm h}$ 31$^{\rm m}$ 34$^{\rm s}$.14}\\
\ \ \ Declination (J2000) 
   & \multicolumn{2}{c}{18$^{\circ}$ 08$\arcmin$ 05\farcs1}\\
Primary Beam HPBW& \multicolumn{2}{c}{$\sim$ 40$\arcsec$}\\
Synthesized Beam HPBW& 
  \multicolumn{2}{c}{3\farcs2 $\times$ 2\farcs0 (P.A. $-60^{\circ}$)}\\
Conversion Factor & \multicolumn{2}{c}{1 (Jy beam$^{-1}$) = 1.6 (K)}\\
Frequency Resolution& \multicolumn{2}{c}{203.125 kHz $\sim$0.178 km s$^{-1}$}\\
Bandwidth& 82 MHz $\times$ 4 & 82 MHz $\times$ 8 \\
Gain Calibrator& \multicolumn{2}{c}{quasar 0423-013} \\
Flux of the Gain Calibrator & 7.8 Jy &  7.2 Jy \\
Passband Calibrator& Saturn& Jupiter \\
Flux Calibrator& Uranus &  Callisto \\
System Temperature (DSB) & \multicolumn{2}{c}{$\sim$350 K}\\
rms noise level (continuum)& \multicolumn{2}{c}{0.02 Jy beam$^{-1}$}\\
rms noise level (line)& \multicolumn{2}{c}{1.2 Jy beam$^{-1}$ / 203.125 kHz}\\
\enddata
\end{deluxetable}

\end{document}